\documentclass{ctr}

\usepackage{ctrfont}
\usepackage{natbib}

\usepackage{url}
\usepackage{amsmath}
\usepackage{amssymb}

\usepackage{graphicx}
\usepackage{psfrag}

\title{Wall-modeled large-eddy simulation of non-equilibrium turbulent boundary layers}
\shorttitle{WMLES of non-equilibrium turbulent boundary layers}
\author{M. Cho, G.~I. Park\footnote{University of Pennsylvania, Philadelphia PA}, A. Lozano-Dur\'{a}n \and P. Moin}
\shortauthor{Cho et al.}

\begin{document}


\maketitle

\section{Introduction} 




Large-eddy simulation (LES) has become an essential tool for both
fundamental studies and real-world engineering applications. However,
industrial use of LES has been hampered by its prohibitive grid-point
requirements near the wall. This limitation motivates the need for
wall models to perform LES at a reduced cost by modeling small-scale
near-wall eddies, while resolving large-scale eddies in the outer
region \citep{Bose2018, Chapman1979, Choi2012}. Although wall-modeled
LES (WMLES) has emerged as a viable alternative to the computationally
more expensive wall-resolved LES \citep{Slotnick2014}, the performance
of wall models in non-equilibrium three-dimensional flows has not yet
been carefully assessed. This assessment is particularly important as
the most widely used wall models to date are built on equilibrium
assumptions.

An example in which non-equilibrium effects emerge are
three-dimensional turbulent boundary layers (3DTBLs). 3DTBLs have
skewed mean velocity profiles, which can be caused by the lateral
motion of the walls or the spanwise pressure gradient imposed by the
bounding geometry. These 3DTBLs exist in a variety of practical
problems, such as the bow and stern regions of a ship, curved ducts,
and turbomachinery, among others. From the point of view of WMLES,
3DTBLs have peculiar features that are challenging to model, such as
the misalignment of the Reynolds shear stress and the mean shear
stress \citep{Cho2018}.

Recently, WMLES of a temporally-developing 3DTBL in a channel flow
with a sudden imposition of spanwise pressure gradient was conducted
by \cite{Giometto2017} and \cite{Lozano2020}. The authors considered
three wall models approaches: an ordinary-differential-equation
(ODE)-based equilibrium wall model \citep{Kawai2012}, an integral
non-equilibrium wall model \citep{Yang2015}, and a partial-
differential-equation-based non-equilibrium wall model
\citep{Park2014}. The performance of the wall models was assessed in
the transient channel flow at $Re_{\tau}=u_{\tau}h/\nu\approx 1000$,
where $u_{\tau}$ is the wall-shear velocity, $h$ is the channel
half-height, and $\nu$ is the kinematic viscosity. \cite{Lozano2020}
quantified the accuracy in predicting the magnitude and direction of
the wall stress as the complexity of the wall model increases, and
concluded that equilibrium models provide the best trade-off between
accuracy and cost. A slip-velocity model \citep{Bae2018} showed
similar performance to the ODE based equilibrium wall model. In
addition to this flow configuration, \cite{Cho2018} examined the
performance of the ODE based equilibrium wall model for a
spatially-developing 3DTBL inside a bent square duct at
$Re_{\tau}=1200-2400$, following the experiment of
\cite{Schwarz1994}. The authors reported a fairly good prediction of
velocity and pressure distributions, with a measurable discrepancy of
the crossflow angles in the bend region. However, only one WMLES mesh
was considered without a grid convergence study.

Another important example of non-equilibrium flows can be found in
separated turbulent boundary layers. These 3D separated flows are
common in engineering applications, and their accurate prediction is
critical during the designing process. Nonetheless, it is unclear if
separated regions should be tackled with WMLES, and the question of
whether a wall model is needed in the separated region remains
unknown. The recent review by \cite{Bose2018} has also pointed out
that the use of the no-slip boundary condition might be justified in
the separated region since there the flow is governed by slow and
large-scale eddies. In that case, it remains to establish how to
switch between the no-slip and wall model boundary conditions.

In this brief, we first report the performance of the ODE-based
equilibrium wall model in a spatially-developing 3DTBL inside a bent
square duct, which is complemented with a mesh convergence study in
Section \ref{duct}. In Section \ref{bump}, we assess the ODE-based
equilibrium wall model for LES predictions of 3D separated
flows. Recent experimental studies by \cite{Ching2018a} on flows over
a wall-mounted skewed bump with 3D flow separations are chosen for
comparison purposes. Concluding remarks are offered in Section
\ref{conclusions}.


\section{WMLES of a spatially-developing 3DTBL in a bent square duct}\label{duct}

\subsection{Computational details}\label{duct_comp}

\begin{figure}
\begin{center}
\includegraphics[width=0.7\textwidth]{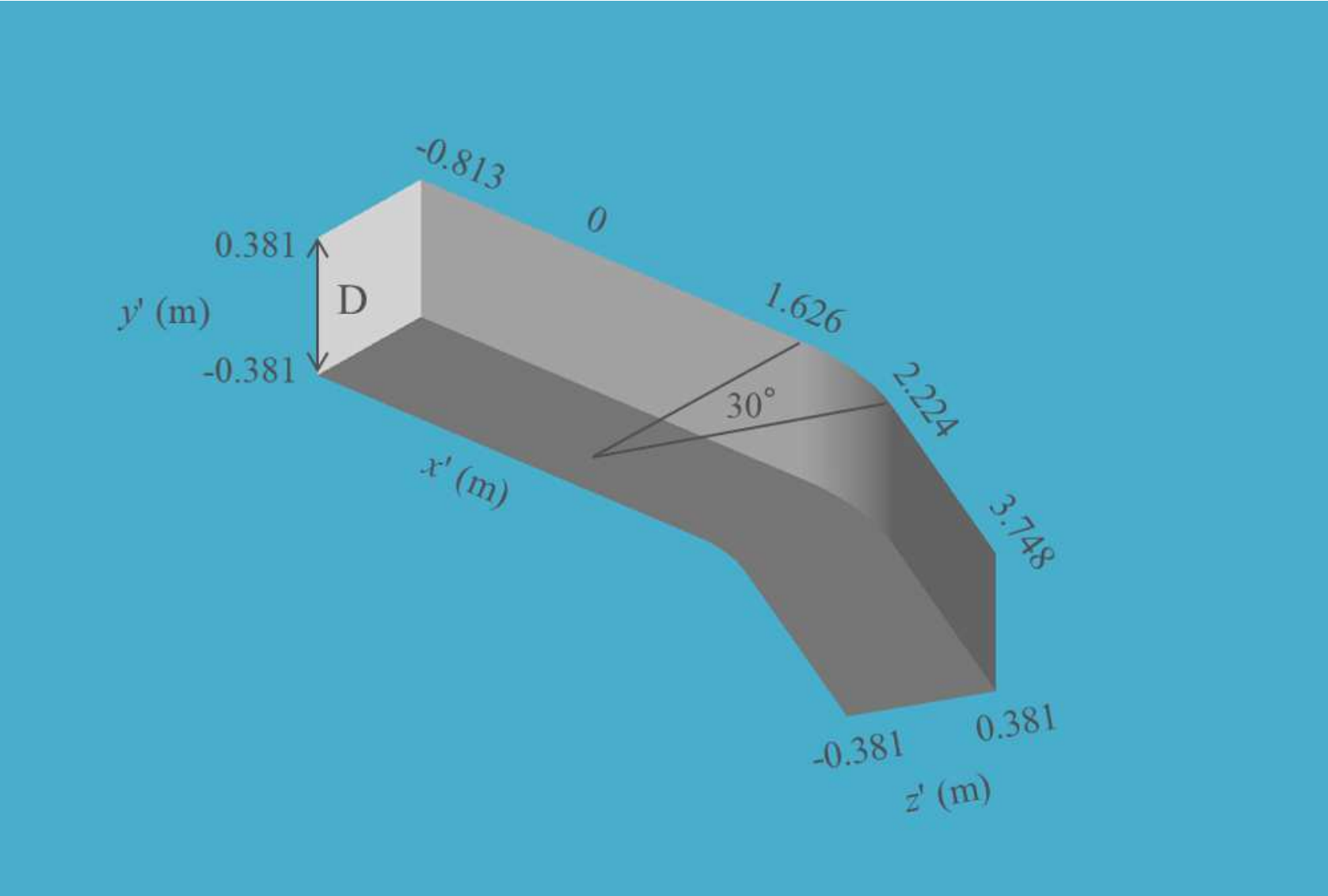}
\caption{Flow configuration of the bent square duct. Here, the coordinate system $(x^\prime,y^\prime,z^\prime)$ is aligned with the local duct centerline. \label{fig1-1}}
\end{center}
\end{figure}

\begin{table}
\caption{\label{table1} Case setup for the bent square duct}
\centering
\begin{tabular}{lcccc}
\hline
Case& $\Delta s$ (m)& $N_{CV}$& $N_{\delta}$& $N_{\delta,\textmd{bend}}$\\\hline
D1& 0.0025--0.01& 30,400,000& 8--11& 9\\
D2& 0.0025--0.01& 38,000,000& 8--15& 13--15\\
D3& 0.00125--0.01& 76,400,000& 8--24& 24\\
\hline
\end{tabular}
\end{table}

Figure \ref{fig1-1} shows the flow configuration of the bent square
duct. The square duct has a 0.762m $\times$ 0.762m cross-section and a
streamwise length of 4.561 m. A spatially-developing 3DTBL is
generated by a 30$^\circ$ bend that imposes a cross-stream pressure
gradient. The surface streamlines are deflected by up to 22$^\circ$
relative to the centerline velocity vector. Then, the developed 3DTBL
gradually recovers to a two-dimensional flow downstream of the
bend. Following the experiment by \cite{Schwarz1994}, two coordinate
systems are defined in the present study (see Figure
\ref{fig1-1}). The coordinate system $(x,y,z)$ is aligned with the
upstream section of the bend, where $x$, $y$, and $z$ denote the
streamwise, wall-normal, and spanwise directions, respectively. The
other coordinate system $(x^\prime,y^\prime,z^\prime)$ is aligned with
the local duct centerline. $(U,V,W)$ are the corresponding mean
velocity components for both coordinate systems.

WMLES is conducted using the code CharLES with Voronoi mesh generator
\citep{Bres2018}. CharLES solves the compressible LES equations with
constant-coefficient Vreman model as the SGS model. The unstructured
Voronoi mesh generator, based on a hexagonal close packed (HCP)
point-seeding method, can automatically build high-quality meshes for
arbitrarily complex geometries with minimal user input. First, a
surface geometry of the square duct is needed to describe the
computational domain, as shown in Figure \ref{fig1-1}. Second, the
user specifies the coarsest grid resolution of the uniformly seeded
HCP points, $\Delta s_\textrm{max}$. For the present WMLES, $\Delta
s_\textrm{max}$ is set to $0.01$ m and three different mesh
refinements are considered, as shown in Table \ref{table1}. For case
D1, the meshes are refined in the near-wall region so that the number
of grid cells across the local boundary layer thickness ($N_{\delta}$)
ranges from 8 to 11 along the streamwise direction. The minimum cell
size in the wall units is $\Delta s_\textrm{min}^+=\Delta
s_\textrm{min} u_{\tau} / \nu=140$, and 30.4 million control volumes
are used in total. Case D2 has additional grid refinement, such that
the number of control volumes across the boundary layer thickness
within the bend section ($N_{\delta,\textmd{bend}}$) increases from 9
(case D1) to 13 to 15 (case D2) using 38 million control volumes. In
the finest grid resolution (case D3), $\Delta s_\textrm{min}$ is
further reduced to $0.00125$ m, resulting in
$N_{\delta,\textmd{bend}}=24$ using 76.4 million control volumes.

The wall-shear stress from the ODE-based equilibrium wall model
\citep{Kawai2012} is imposed as the boundary condition at the bottom,
top, and side walls. The walls are assumed to be isothermal. At the
matching location of the equilibrium wall model, temporally-filtered
LES data are provided to the wall model as suggested by
\cite{Yang2017}. The characteristic-based non-reflective outflow
boundary condition is imposed at the outflow plane
\citep{Poinsot1992}. For the inflow boundary condition, a synthetic
turbulence boundary condition is imposed to provide a realistic
turbulent inflow condition that matches the experiment in the upstream
section of the bend (see Figure \ref{fig1-2}). The Reynolds number
based on the duct side length ($D=0.762$ m) and the inlet freestream
velocity (26.5 m/s) is 1,400,000. The Reynolds number based on the
local momentum thickness and the freestream velocity ranges from 4,000
to 9,000 ($Re_{\tau}=1,200-2,400$).

\subsection{Results and discussion}

\begin{figure}
\begin{center}
\includegraphics[width=1.0\textwidth]{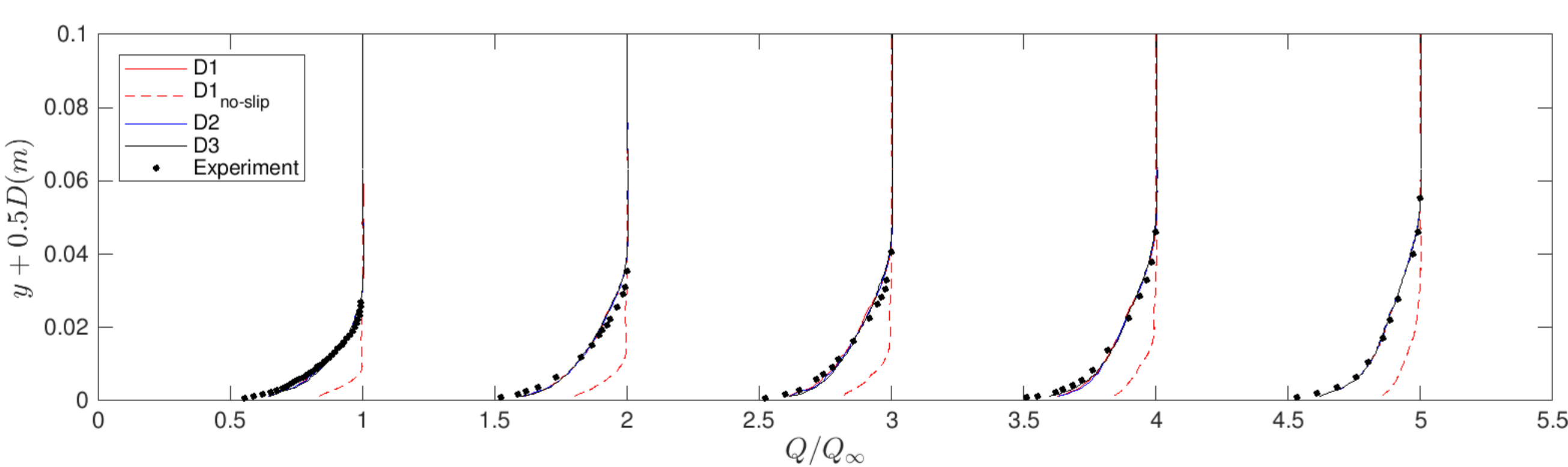}
\caption{Mean velocity profiles at $x^\prime=0.978$ m (upstream
  section of the bend), $x^\prime=1.775$ m and $2.075$ m (inside the
  bend), and $x^\prime=2.415$ m and $2.948$ m (downstream section of
  the bend). Here, the profiles at $x^\prime=1,775, 2.075, 2.415,$ and
  $2.948$ m are shifted by $1,2,3,$ and $4$ on the horizontal axis,
  respectively. Lines, present WMLES (cases D1, D2, D3); circles,
  experiment \citep{Schwarz1994}; dashed lines, no-slip LES on the
  case D1 mesh (without wall model).\label{fig1-2}}
\end{center}
\end{figure}

Figure \ref{fig1-2} shows the profile of the mean velocity magnitude
$(Q=\sqrt{U^2+W^2})$ normalized with the freestream value
($Q_{\infty}$) as a function of the wall-normal distance at various
streamwise locations. The velocity profiles from three grids (i.e.,
cases D1, D2, and D3) are almost on top of each other at each
streamwise location, which indicates that the WMLES solution has
reached grid convergence. For comparison, the results of the no-slip
LES (without wall model) using the same coarse mesh for case D1 are
also included in the figure.

In the upstream section of the bend at $x^\prime=0.978$ m, the WMLES
solution reproduces the upstream condition of the reference
experiment. The mean velocity magnitude profiles inside the bend (at
$x^\prime=1.775$ m and $2.075$ m) and downstream section (at
$x^\prime=2.415$ m and $2.948$ m) are also predicted with reasonable
accuracy. On the other hand, the discrepancy between the experimental
data and the result from no-slip LES is remarkable, indicating the
importance of deploying the wall model at the present grid
resolutions.

\begin{figure}
\begin{center}
\includegraphics[width=1.0\textwidth]{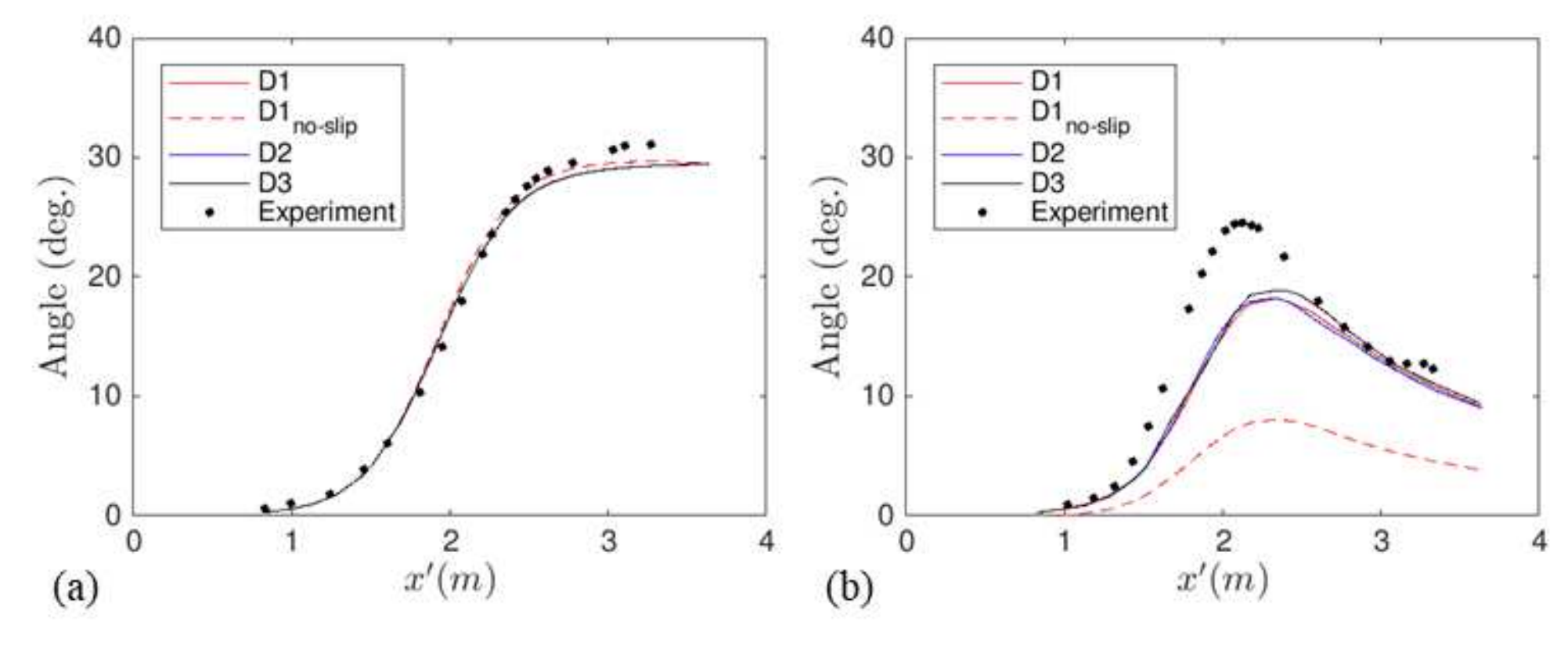}
\caption{Crossflow turning angles: (a) freestream turning angle
  ($\gamma_{\infty}$) distribution along the streamwise direction
  ($x^\prime$); (b) surface crossflow angle relative to freestream
  ($\gamma_s-\gamma_{\infty}$) along the streamwise direction
  ($x^\prime$). Lines, present WMLES (cases D1, D2, D3); circles,
  experiment \citep{Schwarz1994}; dashed lines, no-slip LES on the
  case D1 mesh (without wall model).\label{fig1-3}}
\end{center}
\end{figure}

\begin{figure}
\begin{center}
\includegraphics[width=1.0\textwidth]{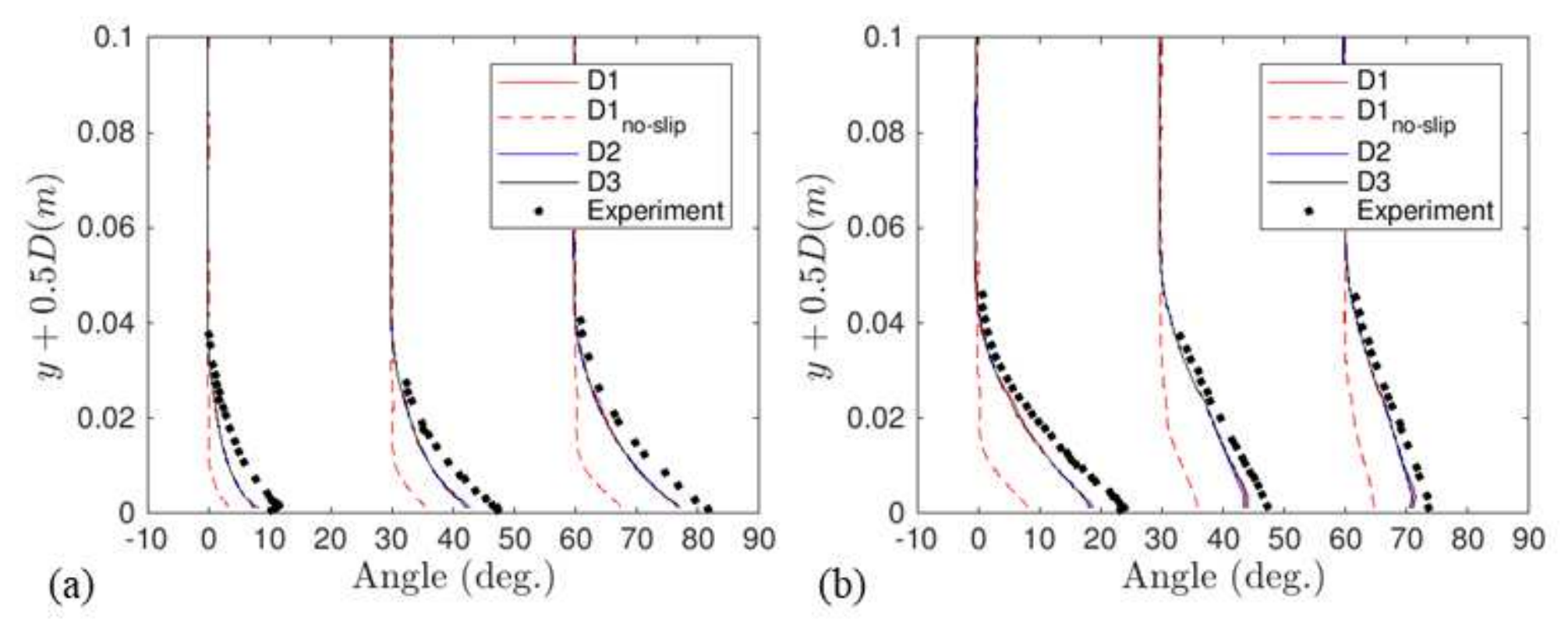}
\caption{Crossflow angle relative to freestream: (a) crossflow inside
  the bend at $x^\prime$(m) = $1.676, 1.875,$ and $2.075$; (b)
  crossflow in the downstream section of the bend at $x^\prime$(m) =
  $2.338, 2.948,$ and $3.329$. Here, the profiles at $x^\prime=1.875$
  and $2.948$ m are shifted by 30$^\circ$ and the profiles at
  $x^\prime=2.075$ and $3.329$ m are shifted by 60$^\circ$ on the
  horizontal axis, respectively. Lines, present WMLES (cases D1, D2,
  D3); circles, experiment \citep{Schwarz1994}; dashed lines, no-slip
  LES on the case D1 mesh (without wall model).\label{fig1-4}}
\end{center}
\end{figure}

Crossflow turning angles are defined as $\gamma=\tan^{-1}W/U$, where
$W$ and $U$ are spanwise and streamwise mean velocity components with
respect to the upstream coordinate ($x,y,z$). Their variations along
the axial and wall-normal directions are represented in Figures
\ref{fig1-3} and \ref{fig1-4}, respectively. Here, $\gamma_{\infty}$
is $\gamma$ at the freestream, while $\gamma_s$ is that at the
surface. $\gamma_{\infty}$ turns from 0$^\circ$ before the bend to
30$^\circ$ after the bend in accordance with the specified geometry,
and the results from both WMLES (cases D1, D2, and D3) and no-slip LES
on the case D1 mesh are almost on top of each other at this freestream
wall-normal location. This is owing to the fact that the inviscid
mechanisms dominate in the core region of the square duct. Also, the
flow turning angles at the freestream wall-normal location
($\gamma_{\infty}$) from both the WMLES and no-slip LES calculations
show excellent agreement with the experiment.

The increase of the surface crossflow in the upstream and bend
sections and its gradual decrease are captured from the current WMLES
for all the cases D1, D2, and D3, but the magnitude of
$\gamma_s-\gamma_{\infty}$ is underpredicted near the bend region. The
result from case D3 is slightly better than for cases D1 and D2, but
the difference is marginal. Here, the maximum value of $\gamma_s$ is
larger than the 30$^\circ$ bend angle, and this shows the effect of
the cross-stream pressure gradient generated by the bend on the
surface streamline direction. Also, the differences between the
current WMLES and experiment are observed within the bend, but these
differences decrease as the 3DTBL inside the bend gradually recovers
to the two-dimensional turbulent boundary layer in the downstream
section.

Figure \ref{fig1-4} shows the crossflow angle relative to the
freestream flow with respect to the wall-normal direction. The
crossflow angle increases until the end of the bend section (Figure
\ref{fig1-4}(a)) and decays in the downstream section (Figure
\ref{fig1-4}(b)). Figure \ref{fig1-4} also shows that the current
WMLES can predict the crossflow development and decay, although the
predicted angles are smaller than those in the experiment. Consistent
with Figure \ref{fig1-3}, the discrepancy between the current WMLES
and the experiment in $\gamma-\gamma_{\infty}$ increases with the
crossflow development and then decreases with its decay. Also, in
Figures \ref{fig1-3} and \ref{fig1-4}, it is shown that no-slip LES
significantly underperforms WMLES.

\section{WMLES of a 3D flow separation behind a skewed bump}\label{bump}

\subsection{Computational details}

\begin{figure}
\begin{center}
\includegraphics[width=0.7\textwidth]{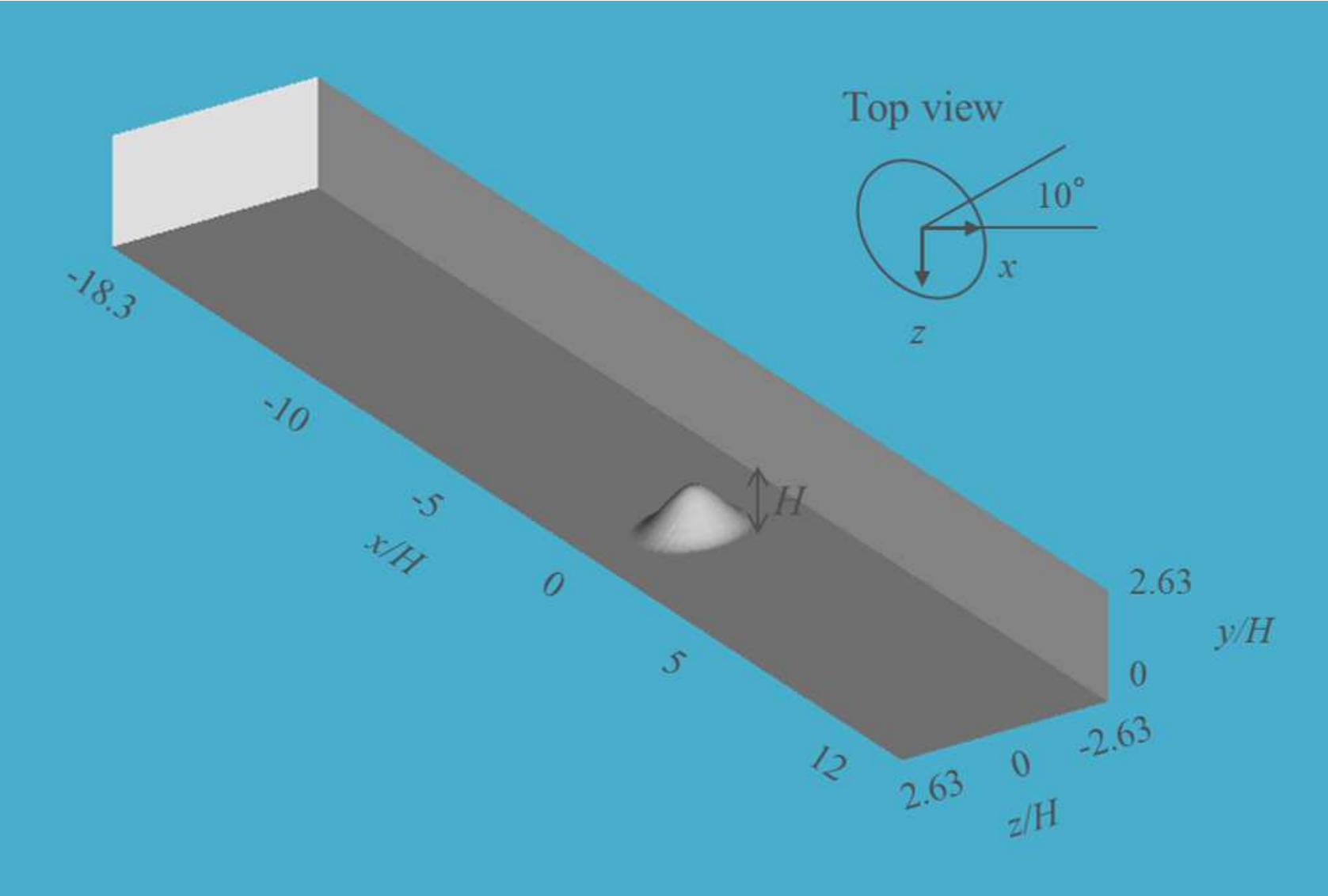}
\caption{Flow configuration of the skewed bump. The coordinate origin
  is located on the bottom wall under the center of the
  bump. \label{fig2-1}}
\end{center}
\end{figure}

\begin{table}
\caption{\label{table2} Case setup for the skewed bump}
\centering
\begin{tabular}{lcccc}
\hline
Case& $\Delta s/H$& $N_{CV}$& $N_{\delta,x/H=-4}$\\\hline
B1& 0.025--0.1&  6,700,000& 20\\
B2& 0.012--0.1& 19,000,000& 20\\
B3& 0.006--0.1& 10,800,000& 20\\
B4& 0.006--0.1& 22,300,000& 20\\
\hline
\end{tabular}
\end{table}

\begin{figure}
\begin{center}
\includegraphics[width=1.0\textwidth]{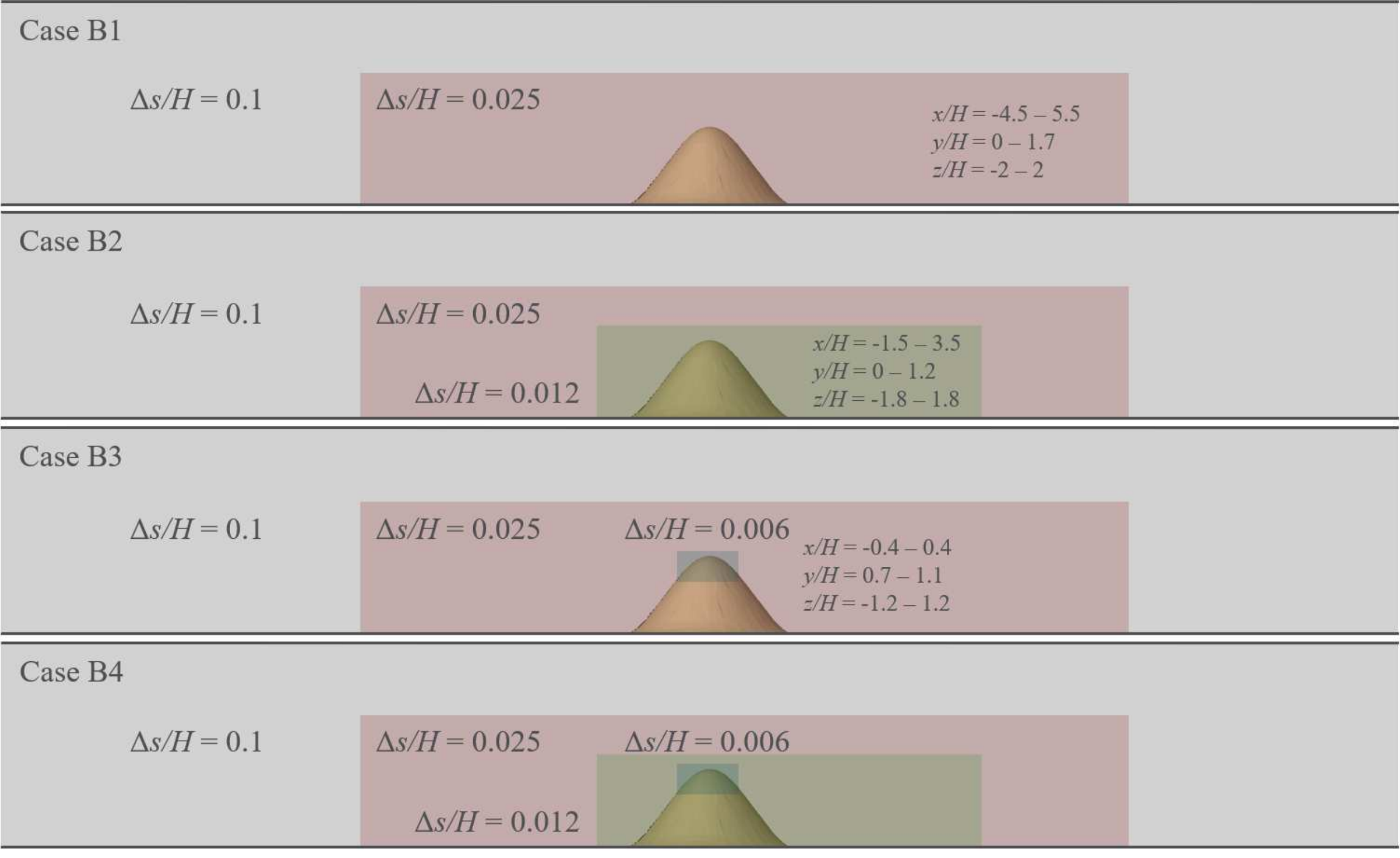}
\caption{Schematic diagram of the mesh refinement. \label{fig2-2}}
\end{center}
\end{figure}

Figure \ref{fig2-1} illustrates the present three-dimensional skewed
bump. The bump is mounted on the bottom wall of the square duct and
its surface is defined by \citep{Ching2018a, Ching2018b, Ching2019}
\begin{equation}
\label{eq1}
y_s = H \left[ 0.5+0.5\cos \left( 2\pi\sqrt{\frac{x_s^2}{b^2}+\frac{z_s^2}{a^2}} \right) \right],
\end{equation}
for $\frac{x_s^2}{b^2}+\frac{z_s^2}{a^2} < \frac{1}{4}$, where $x_s,
y_s$ and $z_s$ denote the surface of the bump in the streamwise,
wall-normal, and spanwise directions, respectively; $H (= 19
\textmd{mm})$ indicates the bump height, $a = 57$ mm, and $b = 42.75$
mm. Here, $x_s$ and $z_s$ are defined with respect to the bump
rotation angle $\theta$ as
\begin{equation}
\label{eq2}
x_s = x \cos \left( \theta \right) + z \sin \left( \theta \right)
\end{equation}
\begin{equation}
\label{eq3}
z_s = x \sin \left( \theta \right) - z \cos \left( \theta \right).
\end{equation}
The bump angle in the present study is at $\theta=10^{\circ}$ and the
simulation results are compared with those of previous experimental
and LES studies \citep{Ching2018a,Ching2019}.  The coordinate origin
is located on the bottom wall under the center of the bump and the
Reynolds number based on the bulk velocity (0.83 m/s), and the bump
height is 16,000.

The same flow solver, CharLES with a Voronoi mesh generator from
Cascade Technologies, is used. The boundary conditions for the skewed
bump are similar to those of the bent square duct (see Section
\ref{duct_comp}). At the inlet, a synthetic turbulence boundary
condition is utilized and the Navier-Stokes characteristic boundary
condition is imposed at the outlet. The ODE-based equilibrium wall
model \citep{Kawai2012} is used to obtain the wall shear stress on the
isothermal bottom, top, and side walls.

The computational domain size is $(L_x \times L_y \times L_z)=(575.7$
mm $\times$ 50 mm $\times$ 100 mm). We consider four different sets of
grid resolutions as shown in Table \ref{table2} and Figure
\ref{fig2-2} to investigate which region needs to be refined to
capture the correct flow physics. For all the cases, the coarsest grid
resolution is set to 0.1$H$, and the meshes are refined around the
bump and the bottom wall ($-4.5H<x<5.5H, 0<y<1.7H, -2H<z<2H$, see
red-colored region in Figure \ref{fig2-2}) such that the grid size in
that region is set to 0.025$H$. This refinement results in 20 grid
cells across the boundary layer thickness at $x/H=-4$. For case B2, an
additional refinement window with the grid size of 0.012$H$ is added
to the case B1 around the bump and the bottom wall ($-1.5H<x<3.5H,
0<y<1.2H, -1.8H<z<1.8H$, see green-colored region in Figure
\ref{fig2-2}). For case B3, the refinement window with the grid size
of 0.006$H$ is added to the case B1 on top of the bump ($-0.4H<x<0.4H,
0.7<y<1.1H, -1.2H<z<1.2H$, see blue-colored region in Figure
\ref{fig2-2}). Lastly for the case B4, both refinement windows used
for each case B2 and B3, respectively, are included into case B1 to
assess grid convergence.

\subsection{Results and discussion}

\begin{figure}\begin{center}
\includegraphics[width=0.7\textwidth]{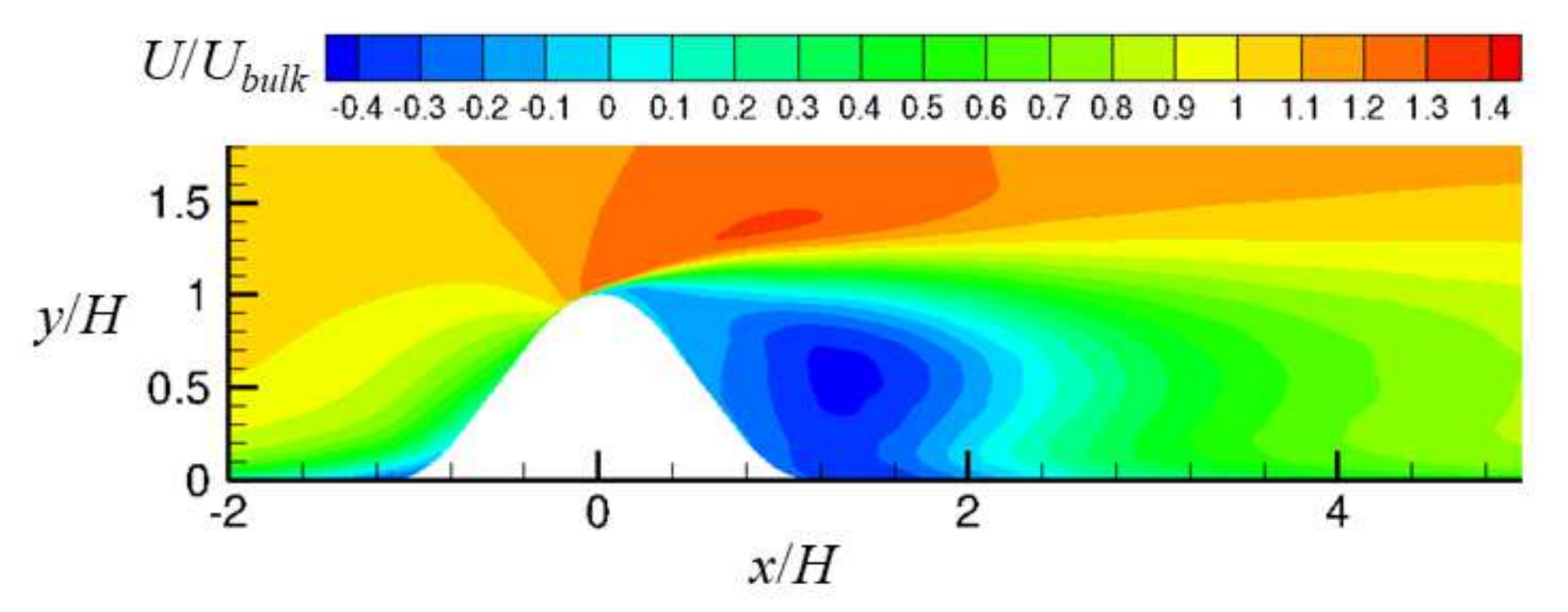}
\caption{Contours of mean streamwise velocity at $z/H=0$ from the case B3.\label{fig2-3}}
\end{center}
\end{figure}

In Figure \ref{fig2-3} shows the mean streamwise velocity contours on
the centerplane (i.e., $z/H=0$) from case B3. The flow separates near
the top of the bump ($x/H \sim 0$) and reattaches around $x/H=2$,
generating a large separation bubble behind the bump. Note that case
B2 is set to contain the main separation bubble within the
green-colored refined region and the case B3 is set to include
additional grid cells around the separation point..

\begin{figure}\begin{center}
\includegraphics[width=0.5\textwidth]{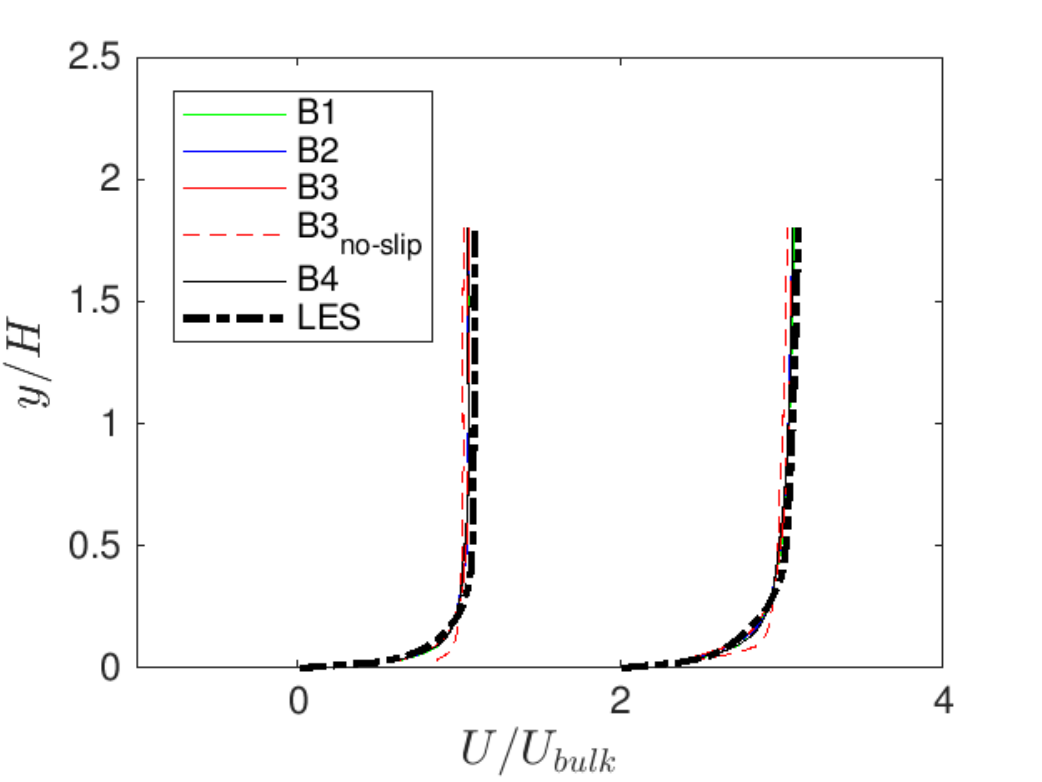}
\caption{Mean velocity profiles in the upstream section of the bump at
  $x/H=-4$ and $-2$. Here, the profile at $x/H=-2$ is shifted by 2 on
  the horizontal axis. Lines, present WMLES (cases B1,B2,B3,B4);
  dashed lines, no-slip LES on the case B3 mesh (without wall model);
  dash-dot lines, LES \citep{Ching2019}.\label{fig2-4}}
\end{center}
\end{figure}

\begin{figure}\begin{center}
\includegraphics[width=0.7\textwidth]{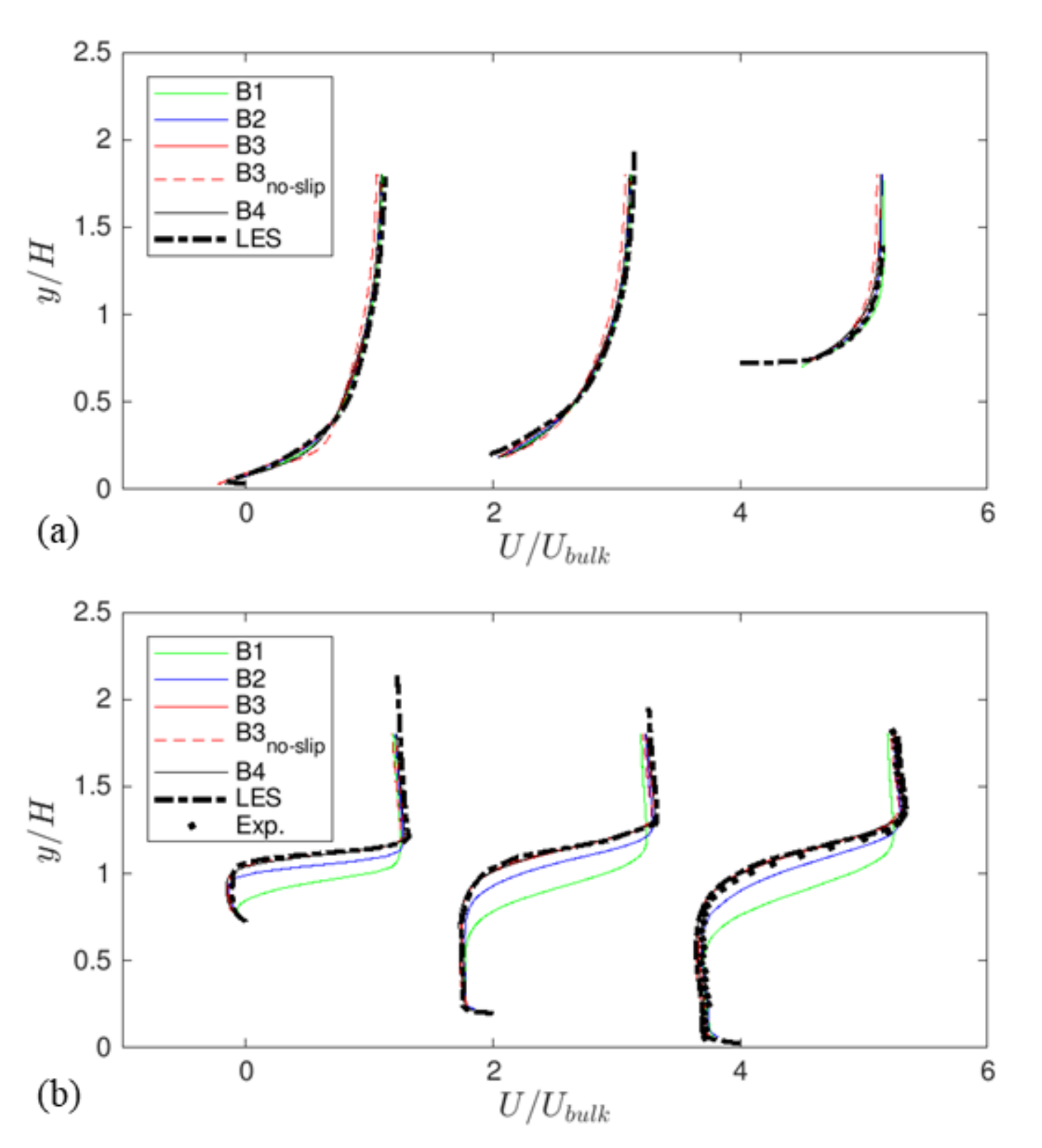}
\caption{Mean velocity profiles over the bump: (a) windward side at
  $x/H=-1,-0.8,$ and $-0.4$; (b) leeward side at $x/H=0.4,0.8,$ and
  $1$. Here, the profiles at $x/H=-0.8$ and $0.8$ are shifted by 2 and
  the profiles at $x/H=-0.4$ and $1$ are shifted by 4 on the
  horizontal axis, respectively. Lines, present WMLES (cases
  B1,B2,B3,B4); dashed lines, no-slip LES on the case B3 mesh (without
  wall model); dash-dot lines, LES \citep{Ching2019}; circles,
  experiment \citep{Ching2018a}.\label{fig2-5}}
\end{center}
\end{figure}

\begin{figure}\begin{center}
\includegraphics[width=0.5\textwidth]{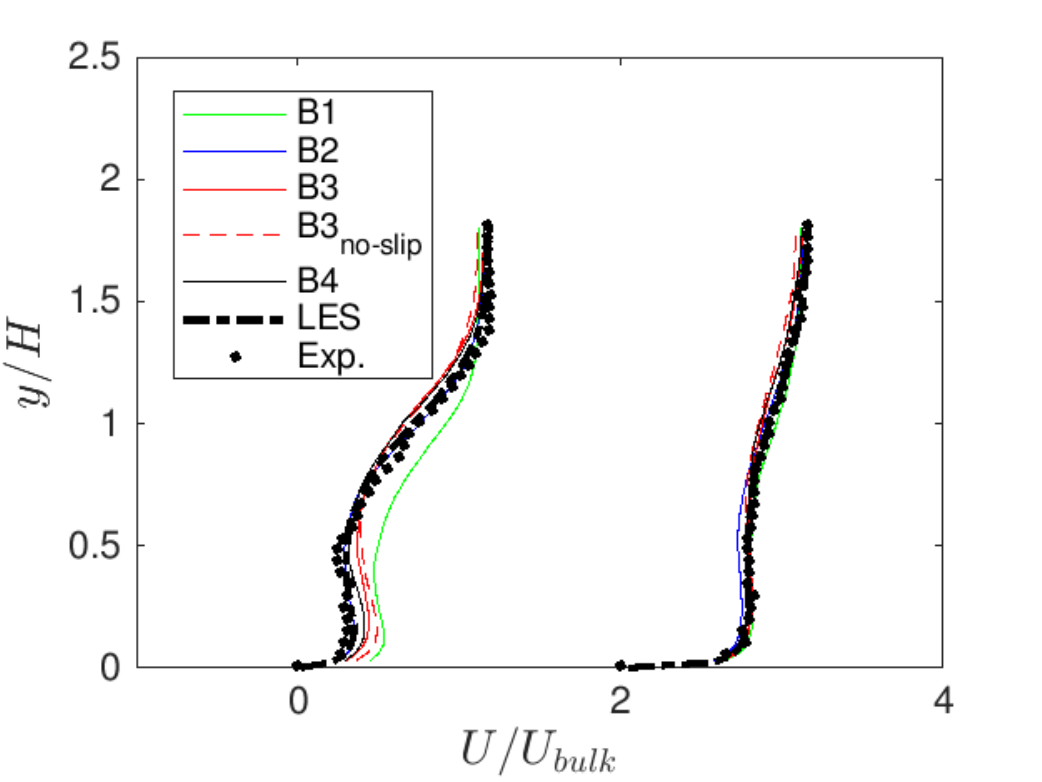}
\caption{Mean velocity profiles in the downstream section of the bump
  at $x/H=3$ and $5$. Here, the profile at $x/H=5$ is shifted by 2 on
  the horizontal axis. Lines, present WMLES (cases B1,B2,B3,B4);
  dashed lines, no-slip LES on the case B3 mesh (without wall model);
  dash-dot lines, LES \citep{Ching2019}; circles, experiment
  \citep{Ching2018a}.\label{fig2-6}}
\end{center}
\end{figure}

\begin{figure}\begin{center}
\includegraphics[width=0.9\textwidth]{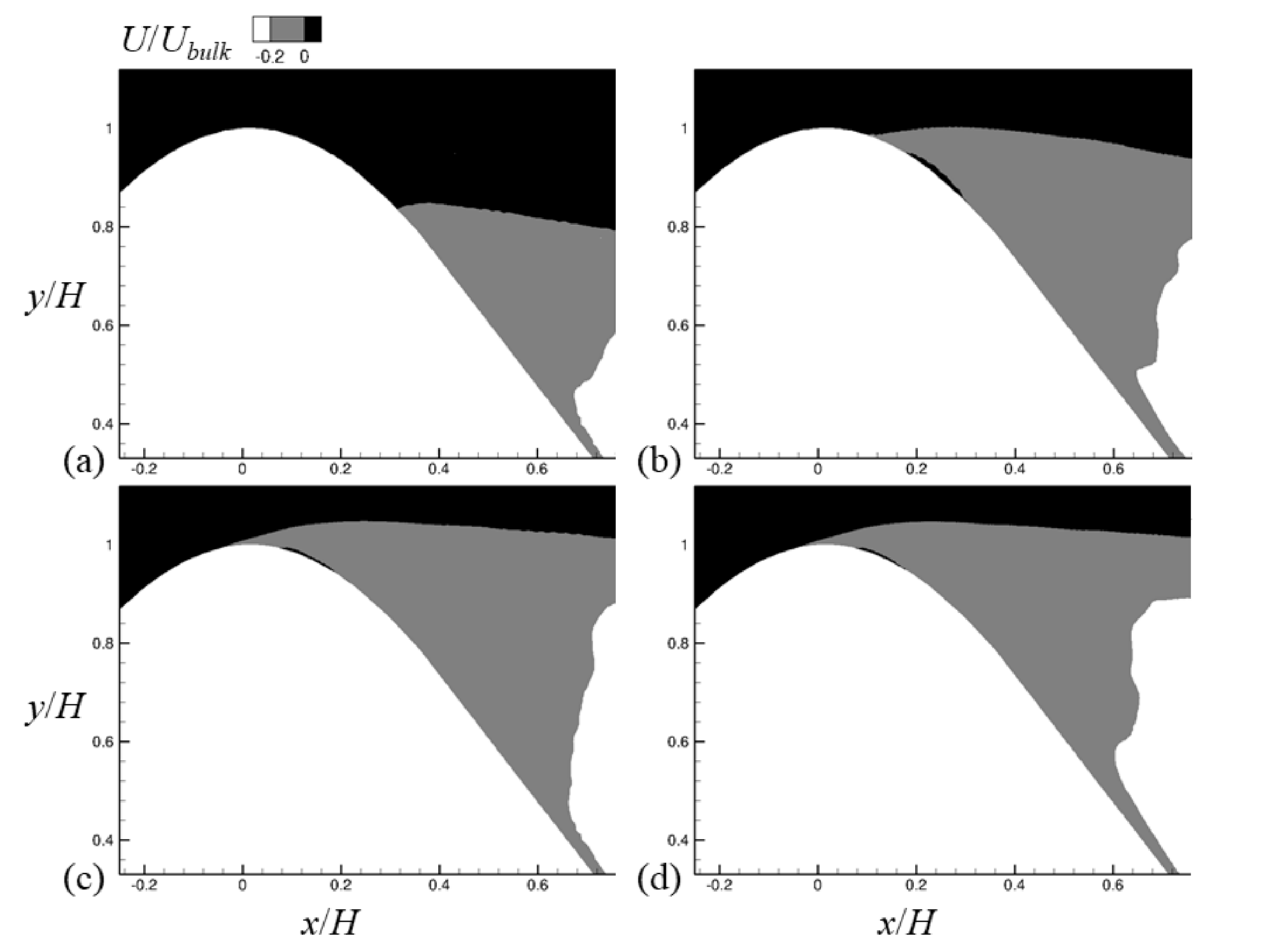}
\caption{Contours of mean streamwise velocity at $z/H=0$ showing the
  separation point and the secondary separation bubble: (a) case B1;
  (b) case B2; (c) case B3; (d) case B4.\label{fig2-6a}}
\end{center}
\end{figure}

Figures \ref{fig2-4}, \ref{fig2-5}, and \ref{fig2-6} show the mean
streamwise velocity profiles at various streamwise locations from the
present WMLES along with the reference experiment and LES data
\citep{Ching2018a,Ching2019}. Before the flow separates around the top
of the bump (i.e., at the upstream section of the bump (Figure
\ref{fig2-4}) and the windward side of the bump (Figure
\ref{fig2-5}(a))), the present WMLES predicts reasonable velocity
profiles regardless of the mesh distribution. This result indicates
that the mean velocity at the upstream section of the bump at $x/H=-4$
can be generated by using the synthetic turbulence boundary condition
as the inlet boundary condition for all the grid resolutions
considered. On the other hand, the velocity profiles at the leeward
side of the bump (Figure \ref{fig2-5} (b)) and at $x/H=3$ in the
downstream section of the bump (Figure \ref{fig2-6}) are sensitive to
the grid resolution. In case B1, the mean velocity distributions at
the leeward side of the bump (Figure \ref{fig2-5}(b)) and at $x/H=3$
(Figure \ref{fig2-6}) do not agree well with the reference experiment
and LES. These locations are inside the main separation bubble and
right behind the bubble, respectively, where the prediction using
WMLES is the most challenging. However, with the additional grid
refinement (cases B2-B4), the discrepancy between the WMLES and
reference cases is greatly reduced; the profiles from the cases B3 and
B4 show an excellent agreement with the reference LES and
experiments. Note that the prediction of the velocity profile from the
case B3 is improved at $x/H=3$ compared to that of the case B1 even
though the mesh is not refined there. Finally, in the far downstream
of the separation bubble at $x/H=5$, the effect of the 3D separation
is weakened and again the present WMLES can predict reasonable
velocity profiles for all mesh resolutions.

The velocity profiles from the no-slip LES (i.e., without the wall
model) on the case B3 mesh are also indicated as the red dashed lines
in Figures \ref{fig2-4}, \ref{fig2-5}, and \ref{fig2-6}. In the
upstream section of the bump (Figure \ref{fig2-4}), these red dashed
lines deviates slightly from the WMLES and experimental results in the
near-wall region. However, velocity profiles over the bump (Figure
\ref{fig2-5}) predicted from the no-slip LES are almost on top of the
red solid lines, i.e., the equilibrium wall model does not influence
this region. The results from the 3D separating and reattaching flows
are different from the previously reported 2D separating and
reattaching flows \citep{Park2017}, showing that the no-slip LES with
the coarse mesh significantly underperforms the WMLES. \cite{Park2017}
reported the performance of WMLES of the flow over the NASA hump,
where the wall-mounted hump geometry was homogeneous in the spanwise
direction. Here, the flow over the hump separates around the apex of
the hump because of the sharp curvature change. Therefore, the
separation point for the NASA hump was determined by the hump
geometry, unlike the present study. Here, the separation is caused by
the development of an adverse pressure gradient. Consequently, the
grid restriction for the current 3D separating and reattaching flows
is more strict than that for the NASA hump configuration. More
importantly, this result indicates that the use of the equilibrium
wall model in the 3D separated region does not hinder the performance
of WMLES and switching the wall boundary condition from the wall shear
stress boundary condition to the no-slip boundary condition might not
be necessary if a sufficiently dense grid resolution is provided
around the separation point (case B3). Hence, the use of the wall
shear stress boundary condition across the full domain might be
justified in the 3D separated flows similar to the present one. Right
after the separation bubble at $x/H=3$, no-slip LES slightly
underperforms the WMLES, but this deviation becomes marginal at
$x/H=5$ in Figure \ref{fig2-6}.

Figure \ref{fig2-6a} shows a zoomed-in view of the streamwise velocity
contours at $z/H=0$ from cases B1-B4 of the present WMLES. In the
reference LES study \citep{Ching2019}, the separation point on the
centerplane was shown to be located around $x/H=0$. Here, this
separation point is not well predicted from the cases B1 and B2,
whereas that from the cases B3 and B4 is similar to the LES result. In
addition to the large separation bubble, the thin secondary separation
bubble was also observed at the leeside of the bump in the previous
LES study by \cite{Ching2019}. These secondary separation bubbles are
captured in Figure \ref{fig2-6a} in the range of $0.1 < x/H < 0.3$ for
cases B2-B4. However, their size and location do not match accurately
with those from the LES study because this secondary bubble is too
thin to be captured with the coarse mesh resolution used in the
current WMLES.

The statistics from case B3 with 10.8 million control volumes are
comparable to those from the case B4 with 22.3 million control volumes
as well as those from the reference studies. This similarity reveals
that the mesh resolution around the 3D separation point (i.e., the
blue refinement window in Figure \ref{fig2-2}) plays an important role
in the successful prediction of the 3D separated flows using the
equilibrium wall model.

\begin{figure}\begin{center}
\includegraphics[width=0.7\textwidth]{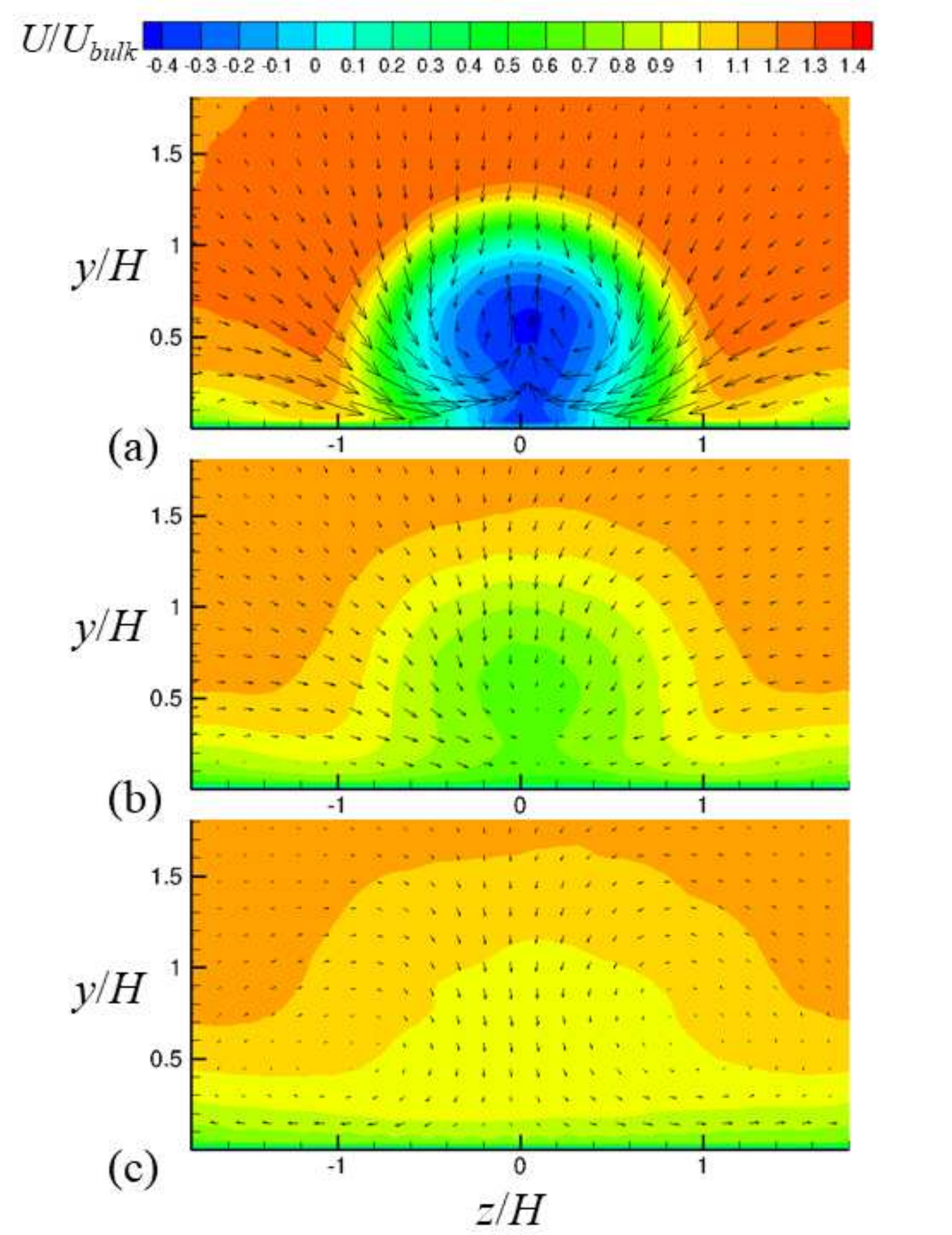}
\caption{Mean streamwise velocity contours with in-plane velocity
  vectors from the case B3: (a) $x/H=1.5$; (b) $x/H=4$; (c)
  $x/H=8$. \label{fig2-7}}
\end{center}
\end{figure}

\begin{figure}\begin{center}
\includegraphics[width=0.7\textwidth]{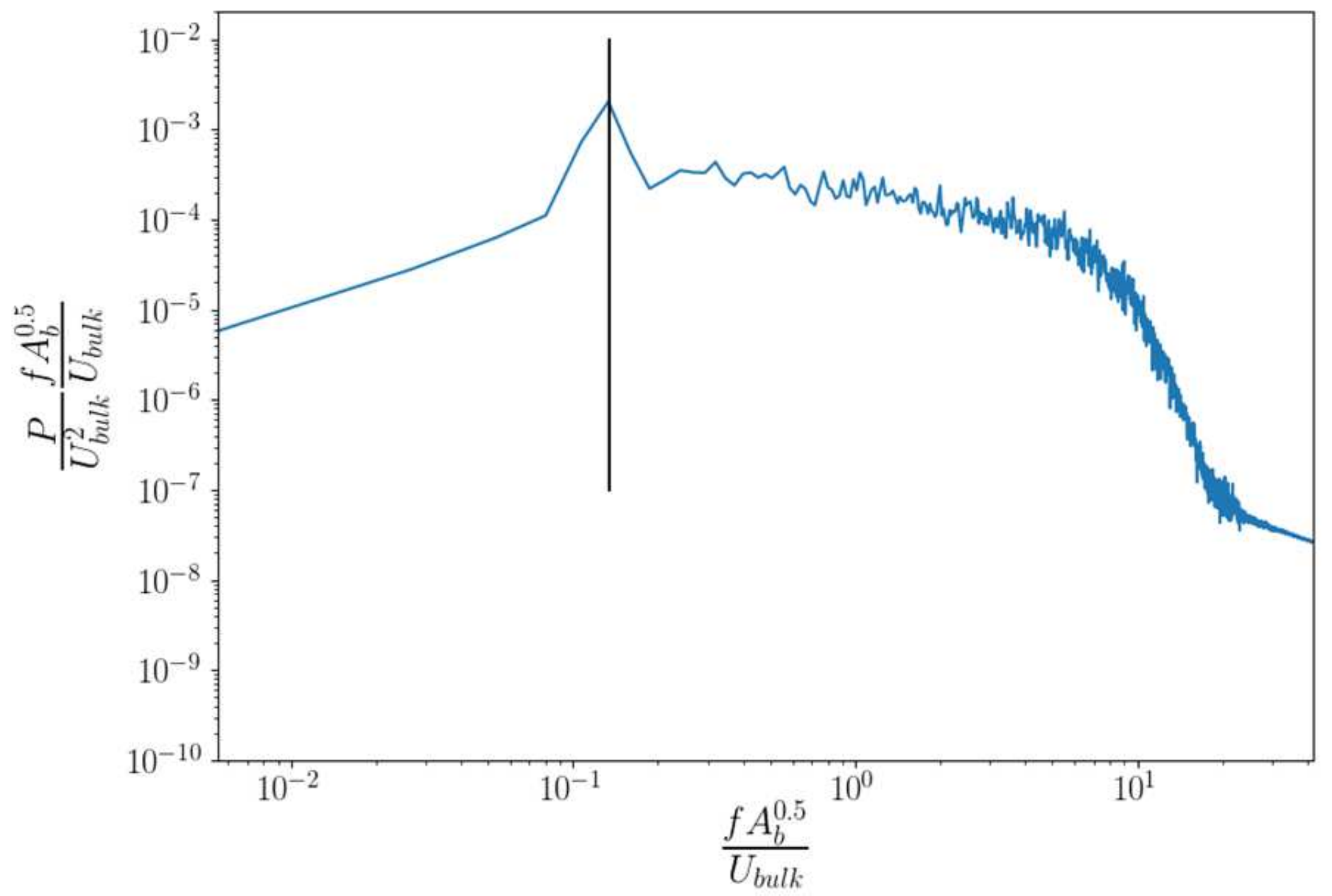}
\caption{Velocity power spectra in the wake region from the case
  B3. The vertical line denotes Strouhal number from LES
  \citep{Ching2019}. \label{fig2-8}}
\end{center}
\end{figure}

Figure \ref{fig2-7} illustrates the vortex structures behind the bump. As
was reported in the previous series of papers
\citep{Ching2018a,Ching2018b,Ching2019}, a common-up vortex pair is
captured right behind the bump (Figure \ref{fig2-7}(a)) which evolves
into a common-down vortex pair in the far downstream (Figure
\ref{fig2-7}(c)). Not only the mean structures, but also the dynamics
of flow structures behind the bump are well captured from the current
WMLES. Figure \ref{fig2-8} shows velocity power spectra from three
probes in the wake region. Following the references
\citep{Ching2018a,Ching2019}, the Strouhal number is defined as
\begin{equation}
\label{eq4}
St = \frac{f \left(A_b\right)^{0.5}}{U_{\mathrm{bulk}}},
\end{equation}
where $f$ is the shedding frequency, $A_b$ is the blockage area, and
$U_{bulk}$ is the bulk velocity. The shedding frequency from the WMLES
matches with the reference LES, indicating that the current WMLES is
able to predict both the statistics and dynamics of the flow over the
skewed bump with reasonable accuracy.

\section{Conclusions}\label{conclusions}

We conducted WMLES to examine the performance of a simple and widely
used ODE-based equilibrium wall model in a spatially-developing 3D TBL
inside a bent square duct \citep{Schwarz1994} and 3D separated flows
behind a skewed bump \citep{Ching2018a,Ching2018b,Ching2019}. From the
square duct simulation, the mean velocity profiles and crossflow
angles in the outer region were predicted with high accuracy for all
the considered mesh resolutions. Some disagreement was observed in the
crossflow angles in the bend region where the non-equilibrium effect
is most significant. Also, the simulation for the wall-mounted skewed
bump showed that this simple ODE-based equilibrium wall model along
with an adequate grid resolution around the 3D separation point
resulted in reasonable predictions of 3D separating and reattaching
flows, including mean velocity distributions, separation bubbles, and
vortex structures in the bump wake.

From the present study, we have demonstrated the potential of the
ODE-based equilibrium wall model for practical LES with relatively
complex flow configurations, despite its low computational complexity
and underlying equilibrium assumptions. It was also shown that the use
of wall shear stress boundary conditions across the whole domain for
3D separated flows (without switching to the no-slip condition) does
not to hinder the performance of WMLES.

\section*{Acknowledgments}

This investigation was funded by ONR Grant No.~N00014-16-S-BA10. MC is
grateful to Dr. David S. Ching for providing the LES data. 

\bibliographystyle{ctr}


\end{document}